\documentclass[fleqn,reprint,groupedaddress,amsmath,amssymb,aps,prx,floatfix,longbibliography]{revtex4-1}
\usepackage{graphicx}

\usepackage{amsmath}
\usepackage{multirow} 
\usepackage{physics} 
\usepackage{gensymb}
\graphicspath{{Figures/}}
\title{TweezerClock}
\date{March 2019}

\begin{document}
\title{Seconds-scale coherence on an optical clock transition in a tweezer array}

\author{Matthew A. Norcia}
\author{Aaron W. Young}
\author{William J. Eckner}
\author{Eric Oelker}
\author{Jun Ye}
\author{Adam M. Kaufman}
\email[E-mail: ]{adam.kaufman@colorado.edu}
\address{JILA, University of Colorado and National Institute of Standards and Technology, and
Department of Physics, University of Colorado, Boulder, Colorado 80309, USA}

\begin{abstract}
	
Coherent control of high-quality-factor optical transitions in atoms has revolutionized precision frequency metrology. Leading optical atomic clocks rely on the interrogation of such transitions in either single ions or ensembles of neutral atoms to stabilize a laser frequency at high precision and accuracy. In addition to absolute time-keeping, the precision and coherence afforded by these transitions has enabled observations of gravitational time dilation on short length-scales, and facilitated applications in quantum information. Here, we demonstrate a new platform for interrogation and control of an optical clock transition based on arrays of individual strontium atoms held within optical tweezers that combines key strengths of these two leading approaches. We report coherence times of 3.4 seconds, record single-ensemble duty cycles up to 96\% through repeated interrogation, and $4.7 \times 10^{-16}(\tau/s)^{-1/2}$ frequency stability commensurate with present-day leading platforms. These results establish optical tweezer arrays, and their associated capacity for microscopic control of neutral atoms, as a powerful tool for coherent control of optical transitions for metrology and quantum information science.

\end{abstract}
%\left(  

\date{\today}
\maketitle

Optical clocks based on neutral atoms and ions achieve exceptional precision and accuracy \cite{mcgrew2018atomic, oelker2019optical, origlia2018towards, ushijima2015cryogenic, Bloom2014, Chou2010a}, with applications to relativistic geodesy \cite{grotti2018geodesy}, tests of relativity \cite{chou2010}, and searches for dark matter \cite{arvanitaki2015searching}. Achieving such performance requires balancing competing desirable features, including a high particle number, isolation of atoms from collisions, insensitivity to motional effects, and high duty-cycle operation~\cite{Ludlow2015, PoliReview}. Here, we introduce a new platform for high-resolution frequency metrology based on arrays of ultracold strontium atoms confined within optical tweezers that realizes a novel combination of these features by providing a scalable platform for isolated atoms that can be interrogated multiple times. By using optical tweezer arrays  --- a proven platform for the controlled creation of entanglement through microscopic control \cite{Bernien2017, Lester2018, wilk2010entanglement, isenhower2010demonstration} --- this work further promises a new path toward combining entanglement enhanced sensitivities with the most precise optical clock transitions \cite{pohl2014}. 

\begin{figure*}%[!htb]
	\includegraphics[width = 6.75in,]{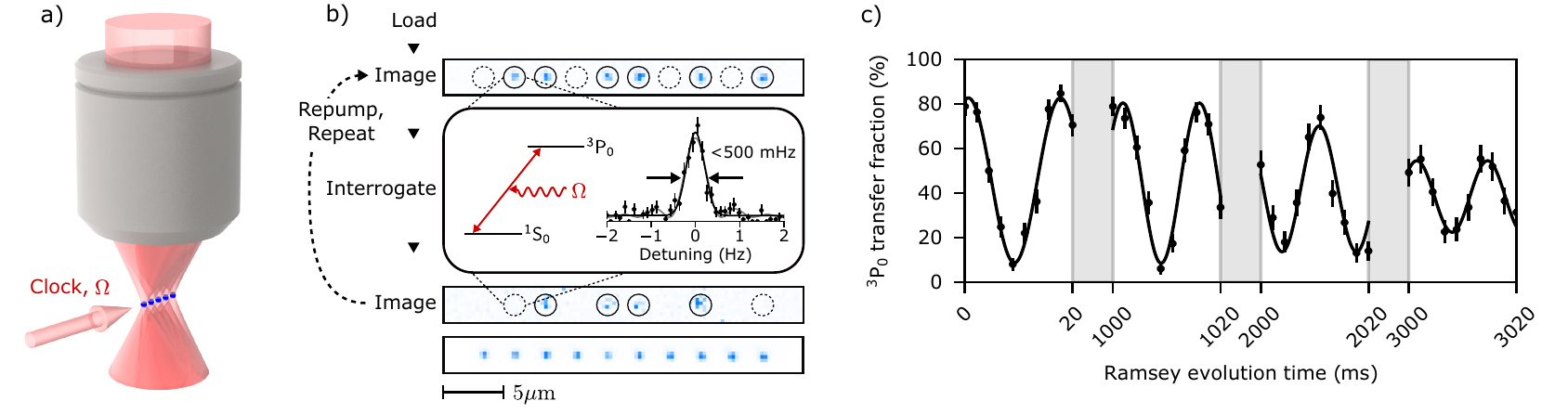}
	\caption{(a) Apparatus for interrogation of $^1$S$_0$ to $^3$P$_0$ ``clock'' transition in tweezer arrays of strontium atoms. Using a high numerical aperture (NA $>0.65$) objective, we project tightly confining optical potentials to trap single strontium atoms. By tuning these traps near the so-called ``magic'' wavelength, we can ensure that the clock transition is minimally sensitive to the local intensity experienced by the atoms. (b) Repeated interrogation of clock transition. Top: image of a single ensemble of atoms loaded into tweezers with $\mathrm{\sim50\%}$ filling. After interrogating the clock transition, excitation of the $^3$P$_0$ state can be inferred from apparent atom loss in a second image. By repumping atoms between interrogation cycles, each ensemble can be interrogated many times before losing atoms. Bottom: image averaged over many such ensembles. Inset: Narrow-line Rabi spectrum of clock transition retrieved without repeated interrogation. Sinc and Gaussian fits shown in gray and black respectively. In this case a 1.5 s probe yields an approximately Fourier-limited Gaussian linewidth of 450(20) mHz. (c) Ramsey spectroscopy in 200~photon recoil energy ($E_R$) deep tweezers, showing a coherence time of 3.4(4) seconds. The frequency of the fringes is set by the differential light shift imposed on the clock transition by the probe beam. These data were taken using the repeated imaging technique outlined in the text. } 
	\label{fig:exp_diagram}
\end{figure*}

The leading platforms for optical clocks --- based on trapped single ions and ensembles of neutral atoms confined within optical lattices --- take distinct approaches in the pursuit of precision and accuracy.  Clocks based on single ions utilize single-particle control and detection to enable high duty-cycle interrogation of an isolated atom \cite{Chou2010a, schmidt2005}. Optical-lattice clocks, on the other hand, typically interrogate thousands of atoms in parallel to reach exceptionally low atom shot noise, enabling frequency measurements with precision below the part-per-$10^{18}$ level within one hour \cite{oelker2019optical}.  

Working with large atom numbers creates challenges associated with laser frequency noise aliasing and inter-atomic collisions, which can be difficult to address simultaneously.  Frequency noise aliasing through the so-called Dick effect arises from dead time between subsequent interrogations of the atoms \cite{Dick1987}.  Because optical lattice clocks typically use relatively shallow trapping potentials and destructive imaging techniques, it is necessary to prepare a new ensemble of atoms between interrogations, exacerbating noise from this effect.  As this noise can be a dominant limitation to the precision of optical-lattice clocks, methods have been developed to overcome dead-time effects by interleaving interrogation of two independent clock ensembles in separate chambers \cite{schioppo2017ultrastable, oelker2019optical}, and techniques for high precision nondestructive detection have been explored \cite{vallet2017noise, norcia2016strong}.  

Interatomic collisions can limit both the precision and accuracy of a clock, but can be mitigated by isolating the atoms from one another.  This has recently been demonstrated in a Fermi-degenerate optical lattice clock, in which a well-defined number of atoms occupies each lattice site~\cite{campbell2017fermi}.  This approach relies on evaporative cooling and a quantum phase transition in the Hubbard model ground state, and therefore typically requires long gaps between clock interrogations, leading to increased sensitivity to noise aliasing. The sub-micron-scale optical lattices essential for such Hubbard-regime physics also limit the current record for atom-light coherence (8 seconds), as dephasing from atomic tunneling must be balanced against decoherence due to scattering from the lattice~\cite{campbell2017fermi,hutson2019engineering}. This suggests that the next frontier in atomic coherence will require new tools that allow control of the atomic spacing~\cite{hutson2019engineering}.

In this work, we demonstrate the core capabilities for a new optical frequency metrology platform based on arrays of individual strontium atoms trapped in optical tweezers (see fig.~\ref{fig:exp_diagram}a). The platform combines the high duty cycle and microscopic control techniques of ion clocks with the scaling capacity inherent to neutral atoms, and allows for a high degree of atomic isolation and coherence.  The single-atom occupancy readily achieved in tweezers eliminates perturbations associated with atomic collisions, while the low temperatures and relatively large spatial separation between tweezers can suppress motional and tunneling effects. The tweezer platform also enables rapid, state-selective, non-destructive detection \cite{fuhrmanek2011free, gibbons2011nondestructive, norcia2018microscopic, cooper2018alkaline, saskin2019narrow, covey20192000}, and thus repeated clock interrogation of the same atoms. As we show, the combination of single-atom-resolved spatial detection and seconds-scale coherence further enables observation of sub-Hz scale variations with  sub-micron spatial resolution, which we use to study dephasing phenomena and may find application in studies of weak interactions in neutral atom clocks~\cite{Chang2004, Kramer2016, Marti2018}. These unique experimental conditions allow the duty cycles, stability, and atom-optical coherence reported in this work. 

Combining coherent control of the clock transition with the microscopic control afforded by optical tweezers is not only useful for optical frequency metrology, but is central to several directions in quantum information science broadly.  
Such control is necessary for proposals for quantum gates based on spin-orbital exchange interactions \cite{Hayes2007, Daley2008, pagano2018fast}. Further, single-photon Rydberg transitions from the excited clock state allow access to many-body spin models and gate architectures with fast timescales compared to dissipation rates, which may lead to improvement over analogous schemes with Alkali species~\cite{Bernien2017, pohl2014, wilk2010entanglement, isenhower2010demonstration}.  Rydberg dressing on the clock transition of microscopically controlled atoms also provides a clear path to the generation of entangled states with improved metrological performance~\cite{pohl2014}, a long-standing goal for optical clocks.

\begin{figure*}[!htb]
	\includegraphics[width=4.5in, ]{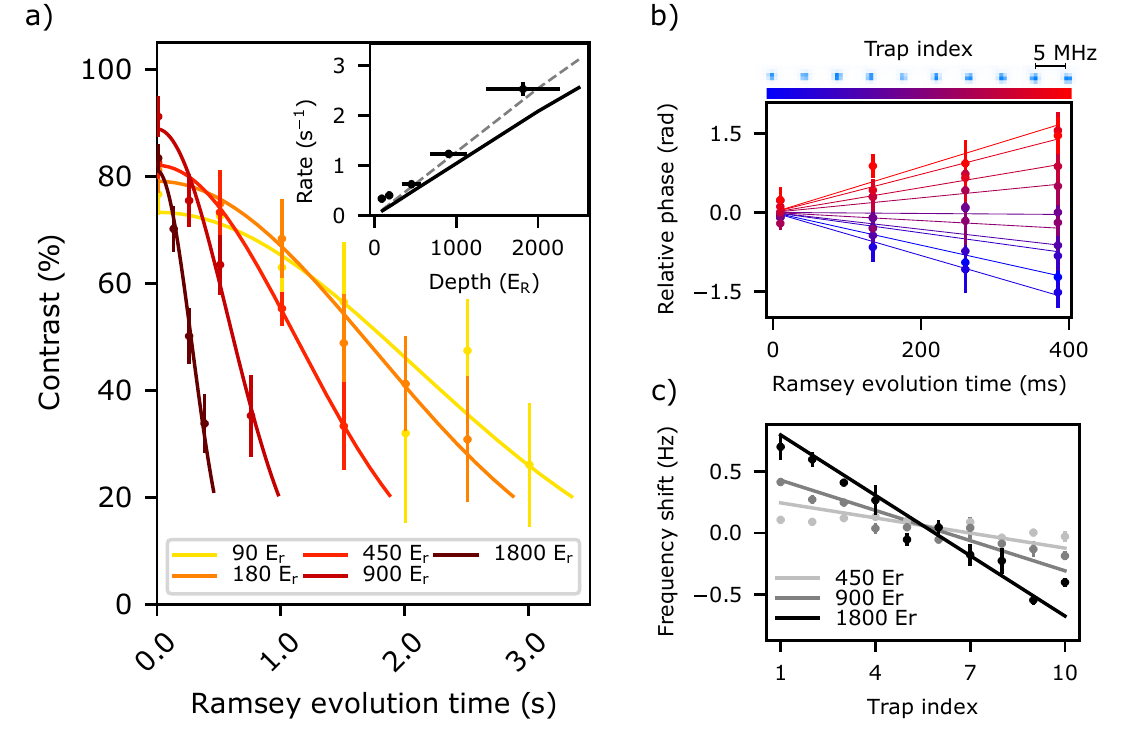}
	\caption{(a) Ramsey contrast as a function of evolution time for different trap depths, showing tweezer-induced contrast decay. Inset: 1/e Gaussian decay rates for Ramsey contrast versus tweezer depth, in units of the tweezer photon recoil energy $E_R$. For deep tweezers, contrast decay rate is proportional to tweezer depth.  This can largely be explained by variations in the frequency of the trapping light across the tweezer array as a result of different RF tones used to generate the tweezer spots, whose expected contribution is represented by the black line. The combined expectation of both this effect and Raman scattering, as inferred from measured depopulation rates from $^3$P$_0$, is illustrated by the gray dashed line~\cite{hutson2019engineering, supplement}. (b) Measured relative Ramsey phase shifts for individual traps in 1800~$E_R$ tweezers.  (c) Inferred relative frequency shifts for individual tweezers at different depths, with predictions based on the known sensitivity to trap detuning \cite{sr87pol, Takano2017}. }
	\label{fig:coherence}
\end{figure*}

Our platform~\cite{norcia2018microscopic} consists of one-dimensional arrays of neutral, bosonic $^{88}$Sr atoms that are tightly confined within optical tweezers, and cooled using three-dimensional sideband cooling (Fig.~\ref{fig:exp_diagram}). With each run of our experiment, individual tweezers have a approximately 50\% probability of being empty or of containing a single atom, where multiple occupancies have been suppressed by light-assisted collisions \cite{Schlosser2001}.  For the current work, we use ten traps, generated by applying ten radio-frequency tones to an acousto-optic deflector.   Imaging is performed with negligible atom loss by simultaneously scattering photons on the broad, $^1$S$_0$ to $^1$P$_1$ transition at 461~nm and cooling on the narrow linewidth $^1$S$_0$ to $^3$P$_1$ transition at 689~nm~\cite{norcia2018microscopic, cooper2018alkaline,covey20192000}.  

We interrogate the $^1$S$_0$ to $^3$P$_0$ optical ``clock" transition, which is induced in our bosonic atoms by applying a magnetic field \cite{taichenachev2006magnetic, origlia2018towards}, using light from a highly stable laser referenced to a crystalline optical cavity \cite{oelker2019optical}. Magnetic field values of 0.14~mT to 2.2~mT and probe intensities ranging from 50~mW/cm$^2$ to 5~W/cm$^2$ are used to generate Rabi frequencies between 0.125~Hz for our narrowest linewidth Rabi spectroscopy and up to 17~Hz for Ramsey spectroscopy \cite{supplement}.  By comparing images of the atomic array taken before and after probing, we infer the excitation to the $^3$P$_0$ state from the apparent loss of atoms in the $^1$S$_0$ ground state. Because the imaging is highly nondestructive \cite{covey20192000, saskin2019narrow}, we can repeat this interrogation cycle many times before preparing a new ensemble of atoms (Fig.~\ref{fig:exp_diagram}b). The tweezer light has a wavelength near 813.4272~nm, where the light-shifts to $^1$S$_0$ and $^3$P$_0$ are nearly equal (the so-called ``magic wavelength" \cite{ye2008}).  Employing a ``magic angle" technique \cite{norcia2018microscopic} allows us to also achieve state-insensitive trapping on the $^1$S$_0$ and $^3$P$_1$ transition at this wavelength, enabling sideband cooling to average phonon numbers of 0.2 along the axis of the clock interrogation. 

These conditions enable coherent atom-light interactions on seconds-long timescales.  Fig.~\ref{fig:exp_diagram}b (inset) shows the inferred excitation probability associated with a 1.5 second laser probe pulse as a function of laser frequency, with the intensity of the probe pulse tuned to maximize transfer probability. The full-width-at-half-maximum of this feature is below 500~mHz, extracted by fitting a Gaussian function to the excitation probability. We further characterize the atom-light coherence through Ramsey spectroscopy, where we scan the duration of the gap between two $\pi/2$ pulses from the clock laser. Because the laser is held at the probe-light-shifted resonance frequency of the clock transition, we observe oscillations in the final transfer fraction occurring at the difference between this frequency and that of the bare clock transition. The contrast of these oscillations persists with a $1/e$ decay time of up to 3.4(4) seconds (fig.~\ref{fig:exp_diagram}c) providing a measure of our atom-light coherence, and a bound on our atom-atom coherence.  

\begin{figure*}[!htb]
	\includegraphics[width=\linewidth, ]{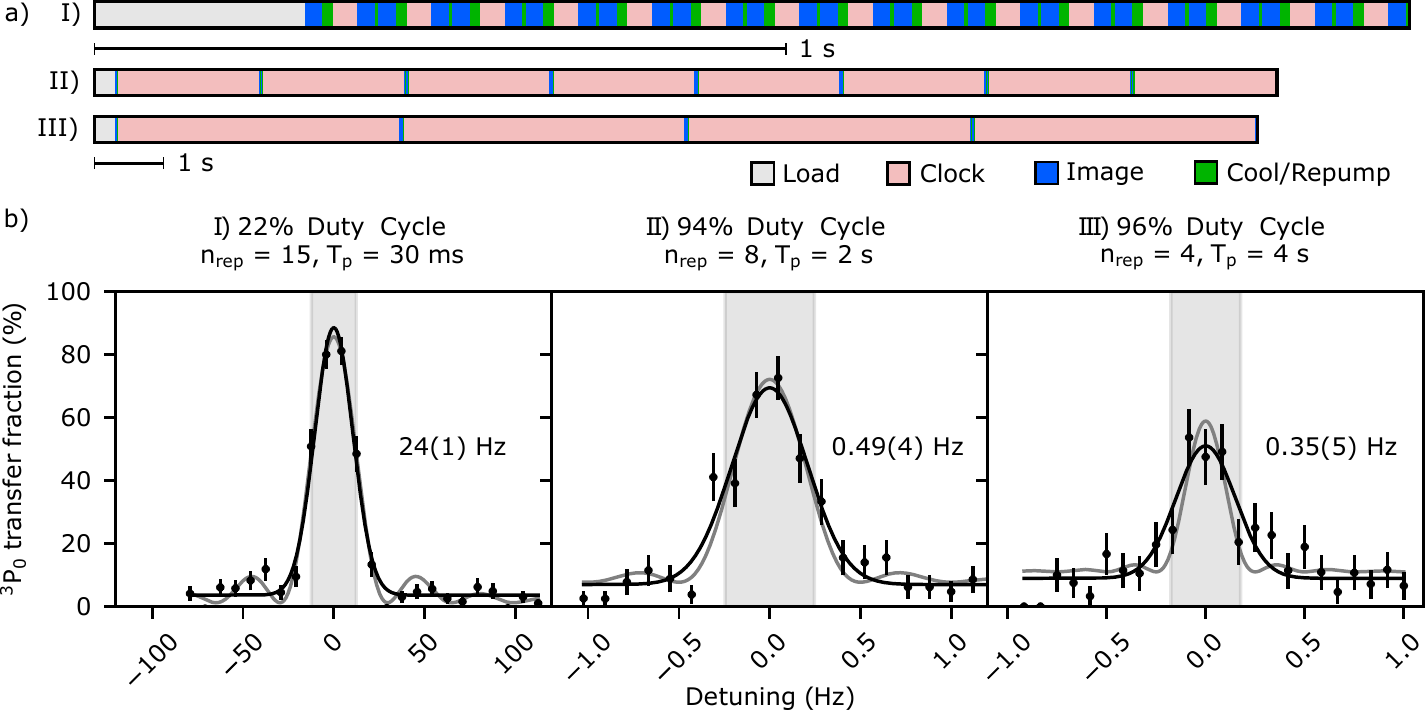}
	\caption{(a) Timing diagram for repeated interrogation of the clock transition for various clock interrogation times ($\mathrm{T_{p}}$) and number of repeated interrogations ($\mathrm{n_{rep}}$): I) $\mathrm{n_{rep} = 15}$, $\mathrm{T_{p} = 30}$ ms; II) $\mathrm{n_{rep} = 8}$, $\mathrm{T_{p} = 2}$ s; and III) $\mathrm{n_{rep} = 4}$, $\mathrm{T_{p} = 4}$ s. These repetition numbers are chosen based on experimental convenience --- high repetition numbers have a more significant impact on duty cycle for short interrogation times, but due to stochastic loading require lengthy data scans for long probe times to achieve uniform statistical uncertainty. 
		(b) Corresponding Rabi spectra for sequences I), II), and III). Sinc fits shown in gray, and Gaussian fits in black. Shaded regions and labels correspond to Gaussian full-width-at-half-maximum.}
	\label{fig:repeated}
\end{figure*}

We observe more rapid decay of Ramsey contrast when operating with deep tweezers, with a rate that depends linearly on the depth of the tweezers (fig.~\ref{fig:coherence}a). This can be largely attributed to slight variations in the optical frequencies of the tweezer light across the different traps, due to the different radio-frequency tones applied to the acousto-optic deflector (adjacent tweezers are separated by 5 MHz).  Because of this frequency difference, all traps cannot be operated at exactly the magic wavelength at the same time.  The predicted magnitude of this effect is illustrated by the black line in the inset of fig.~\ref{fig:coherence}a; the dashed gray line is an expectation that shows the combined effect of this non-magic behavior and the Raman-scattering from the excited state~\cite{hutson2019engineering}.  We use the spatial resolution of our system to study this dephasing effect at a single-atom level in figs.~\ref{fig:coherence}b,c.  For the deepest tweezers used in fig.~\ref{fig:coherence}a, we fit the phase of the Ramsey fringes for each site in the array at different hold times.  We observe a relative shift in this phase between tweezers that increases linearly with hold time and with the distance from the center of the array (fig.~\ref{fig:coherence}b), from which we extract the relative frequency shifts between atoms in different tweezers.  These frequency shifts are plotted in fig.~\ref{fig:coherence}c for selected depths, and agree well with predictions based on the known derivative of the differential polarizability of the clock transition, -15.5(11)~($\mu$Hz/$E_R$)/MHz \cite{sr87pol}.  In order to eliminate this dephasing mechanism in the future, the tweezers could be projected using a digital mirror device or spatial light modulator, for which all traps have the same optical frequency.  

For trap depths below 200 times the photon recoil energy $E_R$ associated with the tweezer light, we find that the improvements in coherence time begin to saturate, indicating the presence of other decoherence mechanisms such as residual path-length fluctuations between the clock laser and the atoms.  Further, we find that the initial contrast of oscillations decreases in shallow tweezers, which could be due to residual atomic motion.  All spectroscopic measurements presented in this work were performed in tweezers with depths of 200 $E_R$ unless otherwise stated.

The alkaline-earth tweezer platform enables non-destructive, state-selective imaging \cite{covey20192000, saskin2019narrow}, which we can use to interrogate the same ensemble of atoms many times and realize high duty-cycle interrogation of the clock transition.  This both improves the rate at which statistical uncertainty can be averaged down and mitigates noise-aliasing of the clock laser \cite{supplement}. Each interrogation cycle involves 90 ms of dead time during which the clock transition is not being interrogated, compared to approximately 300~ms to load new atoms into the tweezers (fig.~\ref{fig:repeated}). For short (30 ms) interrogation pulses, we observe a loss probability of 0.001(1) per cycle. For longer cycle times (up to 8 s), we observe a total loss probability consistent with 0.01 per second for atoms prepared either in the ground or excited states, likely due to collisions with the background gas.

The coherence properties of clock interrogation with repeated imaging are consistent with those when each set of atoms is interrogated only once.  Ramsey contrast curves with 3~second coherence times have been measured using both repeated and single interrogation, as shown in fig.~\ref{fig:exp_diagram}c and fig.~\ref{fig:coherence}a, respectively.  Figure ~\ref{fig:repeated} characterizes this repeated interrogation technique for different clock probe durations ($T_p$) and number of cycles ($n_\mathrm{rep}$) performed using the same ensemble of atoms. 
In fig.~\ref{fig:repeated}b, we use up to 4 second long interrogation pulses to demonstrate both high duty cycles of up to 96\%, as well as clock spectra as narrow as 350~mHz.  

We characterize the short-term stability of the atomic frequency reference by computing the Allan deviation relative to the free-running clock laser.  This measurement is performed using single-interrogation ($n_\mathrm{rep}$=1) Ramsey spectroscopy with evolution times of 501~ms, chosen to bias the signal to the most sensitive part of the Ramsey fringe.  The clock laser was not actively stabilized to the atomic transition, which limits the total length of the dataset to 500 seconds. Under these conditions, we measure a fractional frequency instability of of $4.7 \times 10^{-16}(\tau/s)^{-1/2}$ at short times. As expected, this is dominated by the atomic quantum projection noise associated with the average atom number of 4.8 per trial, while the long-term instability is consistent with known drifts in the frequency of our laser \cite{supplement}.  

\begin{figure}[!htb]
	\includegraphics[width = 3.38in]{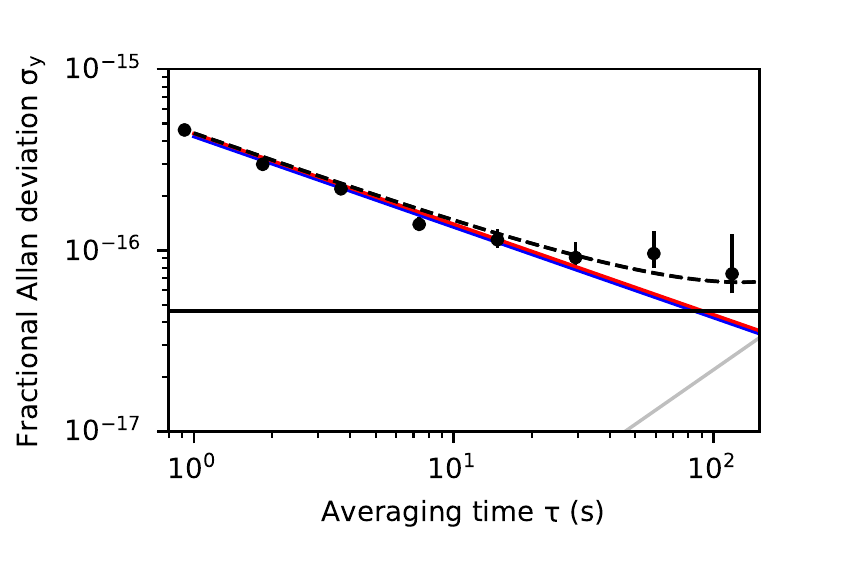}
	\caption{Short-term stability of atomic clock signal relative to interrogation laser, quantified in terms of Allan deviation of fractional frequency fluctuations.  For Ramsey interrogation with 500~ms between pulses, we fit a fractional instability of $4.7 \times 10^{-16}(\tau/s)^{-1/2}$ at short averaging times (red line), consistent with the expected atom shot noise limit after accounting for our measured contrast (blue line).  The flicker noise and known drift of the laser are represented by the black and gray traces respectively, and the predicted combination of atom shot noise, laser noise, and laser drift by the dashed line.   
	} 
	\label{fig:allen}
\end{figure}

Currently, our demonstrated instability falls between that attained with single ions \cite{Chou2010a}, and in optical lattice clocks operating with thousands of atoms \cite{oelker2019optical, mcgrew2018atomic, ushijima2015cryogenic, origlia2018towards}. Achieving stability competitive with the most stable optical lattice clocks ($4.8 \times 10^{-17}(\tau/s)^{-1/2}$\cite{oelker2019optical}) could be achieved with increased atom number -- for example, by using a two-second-long dark time and approximately 150 atoms. Because of our high duty cycle and low-noise laser system, Dick effect noise would contribute only modestly ($16\%$ and $6\%$ of noise variance for individual and sixteen times repeated interrogation respectively). While we demonstrate such coherence times in this work, attaining this atom number would require more power than is available with our current system (here we use 13(1)~mW per tweezer spot at the atoms for cooling and imaging).  However, by performing atomic preparation and imaging in tweezers at a wavelength with higher polarizability, smaller spot size, and higher available power, such as 515 nm, then transferring into 813 nm tweezers for clock interrogation, ensemble sizes of 500 atoms or larger could plausibly be achieved, which would enable stability exceeding the current state of the art.  

In addition to enabling one to take full advantage of the reduced projection noise associated with large numbers of atoms, the low dead times provided by our tweezer platform would allow one to achieve high performance with a less stable clock laser.  This capability could be especially transformative for high-performance portable optical clocks \cite{grotti2018geodesy}, where technical constraints limit the performance of the clock laser.  In this context, our repeated imaging technique would allow for relatively high duty cycles even with short interrogation times, greatly improving the stability associated with both quantum projection noise and laser noise aliasing \cite{supplement}.

For use as an absolute frequency reference, systematic shifts to the clock transition frequency must be carefully considered.  Many of the shifts present in the tweezer platform, particularly environmental perturbations such as blackbody radiation, are identical to those present in optical lattice clocks, and have been studied in detail \cite{Ludlow2015, PoliReview}.  In this work, we use bosonic $^{88}$Sr atoms.  While this choice is not fundamental to our tweezer platform, it does require consideration of different systematic effects than the more commonly used fermions.  Shifts due to atomic collisions are of particular concern in clocks operating with bosonic atoms \cite{origlia2018towards}, but are mitigated in the tweezer platform by single-atom occupancy.  The use of the bosonic isotope requires higher probe intensities and a relatively large applied magnetic field to reach a given Rabi frequency, so second order Zeeman and probe-induced stark shifts can be significant.  For narrowline Rabi spectroscopy, our probe intensities and magnetic fields are similar to those of ref. \cite{origlia2018towards}, where their contributions to clock inaccuracy were bounded to the $10^{-17}$ level.  Unlike conventional optical lattices, tweezer confinement relies on tightly focused beams that can have nonuniform polarization at the location of the atoms.  While such inhomogeneous polarization could complicate magic trapping in fermionic isotopes, its effect is expected to be suppressed in bosons, whose clock states have zero net angular momentum \cite{taichenachev2007optical}.  Finally, in our current setup there are several meters of optical path between the atoms and the surface to which the phase of the interrogation laser is referenced. Fluctuations in this path length lead to Doppler shifts, which can limit stability and in principle accuracy. In analogy to techniques used in optical lattice clocks, this could be addressed by referencing the phase of the interrogation laser to a surface rigidly connected to the microscope objective, as this primarily defines the atoms' locations.  Characterization of these potential systematic effects unique to the tweezer platform, as well as those unforeseen, will be critical to future studies of clock accuracy.

The ability to control and measure the positions and states of individual atoms may enable new opportunities both for extending single-atom coherence, and for utilizing ensembles of many atoms in new ways.  At a single-particle level, the spatial control afforded by the tweezers could allow one to simultaneously suppress tunneling and light-induced decoherence to push coherence times beyond 10 seconds \cite{hutson2019engineering}.  When using many atoms, the microscopic control afforded by the tweezers, combined with interactions introduced using Rydberg dressing \cite{pohl2014}, could facilitate the creation of entangled states that can surpass the limitations set by atomic projection noise, for both fundamental studies of entanglement on an optical clock transition as well as quantum-enhanced high-bandwidth sensors.  Future extensions of the imaging and coherent manipulation techniques demonstrated here, in which sub-ensembles of the atoms are manipulated independently, could also allow for techniques that extend interrogation time beyond the coherence time of the laser \cite{schioppo2017ultrastable} in systems using a single vacuum chamber.

\section{Acknowledgments}

We would like to acknowledge discussions and support from Toby Bothwell, Mark Brown, Julia Cline, Aki Goban, Ross Hutson, Benjamin Johnston, Dhruv Kedar, Colin Kennedy Andrew Ludlow, William Milner, Cindy Regal, John Robinson, Christian Sanner, Juan A. Muniz Silva, Lindsay Sonderhouse, James K. Thompson, and Jeff Thompson. This work was supported by ARO, AFOSR, DARPA, the National Science Foundation Physics Frontier Center at JILA (1734006), and NIST. M.A.N. and E.O. acknowledge support from the NRC research associateship program.

%\bibliography{references}
%\bilbiographystyle{Science.bst}
%

%apsrev4-2.bst 2019-01-14 (MD) hand-edited version of apsrev4-1.bst
%Control: key (0)
%Control: author (8) initials jnrlst
%Control: editor formatted (1) identically to author
%Control: production of article title (0) allowed
%Control: page (0) single
%Control: year (1) truncated
%Control: production of eprint (0) enabled

\clearpage
% \widetext
\begin{center}
\end{center}
%%%%%%%%%% Merge with supplemental materials %%%%%%%%%%
%%%%%%%%%% Prefix a "S" to all equations, figures, tables and reset the counter %%%%%%%%%%
\setcounter{section}{0}
\setcounter{equation}{0}
\setcounter{figure}{0}
\setcounter{table}{0}
\setcounter{page}{1}
\makeatletter
\renewcommand{\theequation}{S\arabic{equation}}
\renewcommand{\thefigure}{S\arabic{figure}}
\renewcommand{\thetable}{S\arabic{table}}
% \renewcommand{\bibnumfmt}[1]{[S#1]}
% \renewcommand{\citenumfont}[1]{S#1}
%%%%%%%%%% Prefix a "S" to all equations, figures, tables and reset the counter %%%%%%%%%%

\section{Supplementary Materials}

\subsection{Materials and Methods}

\noindent \textbf{Tweezer parameters and wavelength. }
For this work, we operate with tweezers at 813.4272~nm, for which the Gaussian waist size is 740(40)~nm.  In all data presented in this work, the depth of the tweezers during spectroscopy are 200~$E_R$ unless otherwise stated, while loading, imaging, light assisted collisions, and sideband cooling are performed in much deeper 3900 to 6000~$E_R$ deep tweezers.  

\begin{figure}[!htb]
	\includegraphics[width = 2.25in]{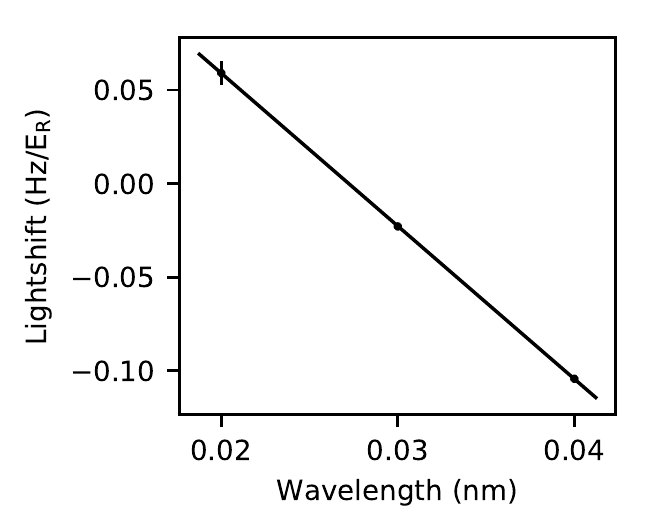}
	\caption{Strength of lightshifts versus tweezer wavelength, with 813.4~nm offset. From a linear fit we extract a magic wavelength of 813.4272(5) nm.}
	\label{fig:MWL}
\end{figure}

We measure the tweezer wavelength at which shifts to the clock transitions disappear -- the magic wavelength -- by performing clock spectroscopy while varying the wavelength and intensity of the tweezer light (fig.\ref{fig:MWL}).  We measure a value of 813.4272(5)~nm using a wavemeter whose calibration we confirm using our spectroscopy-stabilized 689~nm light.  For our typical lattice depths of 200 $E_R$, and combined uncertainty and fluctuations in the frequency of our lattice laser of 250~MHz, we expect errors in our clock spectroscopy of roughly 0.8~Hz.  Such errors could be dramatically reduced by referencing the frequency of our lattice laser to a stable optical cavity or frequency comb, enabling a more precise and stable determination of the magic wavelength.  

Tweezers spots are generated by applying sets of radio frequency tones to an acousto-optic deflector.  The frequency of these tones determines the deflection angle of the beam used to generate the tweezer spots, and in turn the position of the tweezers.  
In this work we operate with a 5~MHz offset between traps (corresponding to 2.5~$ \mathrm{\mu}$m spatial separation). This choice was made out of convenience, and our experiment is compatible with substantially tighter spacings.

\begin{figure*}[!htb]
	\includegraphics[width = \linewidth, ]{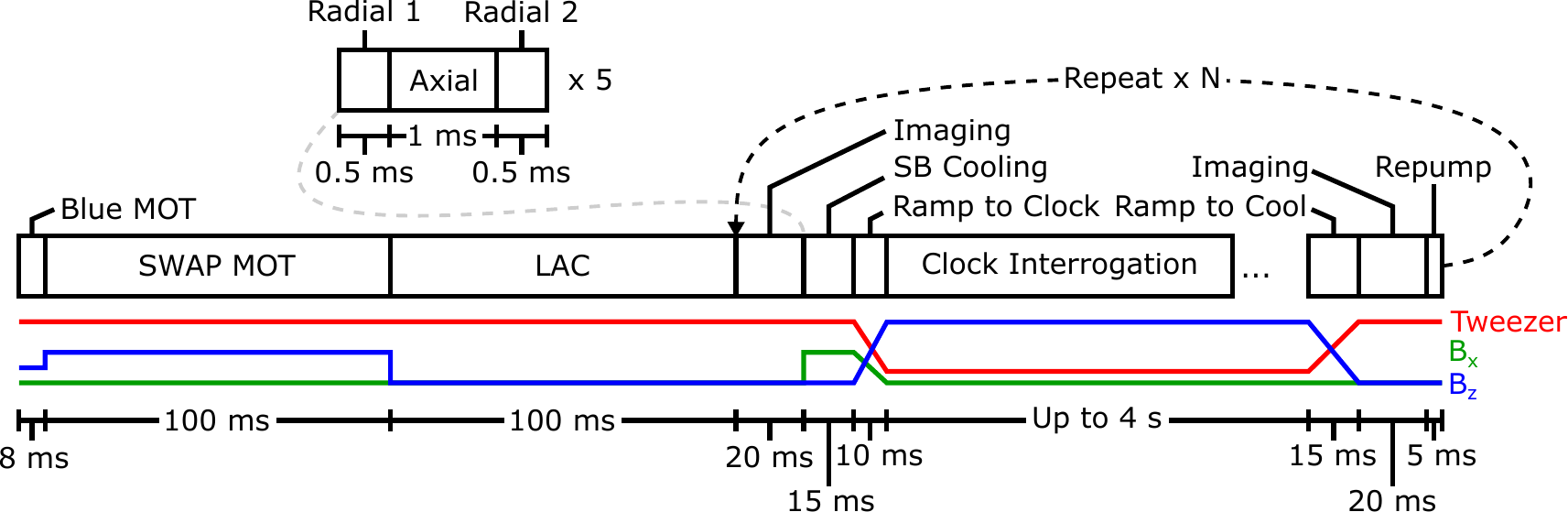}
	\caption{Timing diagram of typical experimental sequence for N repeated interrogations of the clock transition. Callout shows resolved sideband cooling protocol, which involves sequential pulses of sideband cooling light along the longitudinal axis of the tweezers, and along two orthogonal radial directions.}
	\label{fig:sequence}
\end{figure*}

\noindent \textbf{Initial trapping, loading, and imaging. } Our experimental apparatus is described in \cite{norcia2018microscopic}, with modified timings shown in fig.~\ref{fig:sequence}. To project multiply-occupied tweezers to zero or one atoms we induce light-assisted collisions (LAC) by exciting the atoms to the $^3$P$_1$ state. Specifically, at zero field we apply a single beam that is resonant with the AC Stark shifted $^1$S$_0$ to $^3$P$_1$ transition near the bottom of the tweezers (which corresponds to a beam that is 3.5 MHz red-detuned from the free-space $^1$S$_0$ to $^3$P$_1$ transition). This has the additional benefit of simultaneously cooling the atoms in all spatial directions as in \cite{covey20192000}. 

Imaging is performed by using this same cooling method while simultaneously resonantly exciting the $^1$S$_0$ to $^1$P$_1$ transition. Fluorescence from this transition is collected on an electron-multiplying CCD camera. For 20 ms long images in 6000~$E_R$ deep tweezers, this protocol yields less than one atom loss per thousand images, and $99\%$ detection fidelity for ground state atoms.  The use of deep tweezers introduces the possibility of Raman scattering of atoms out of $^3$P$_0$ during imaging, which can lead to incorrect identification of $^3$P$_0$ atoms as being in the ground state.  We measure the depopulation rate of the $^3$P$_0$ state within our tweezers by exciting atoms on the clock transitions, then measuring the fraction of atoms remaining in $^3$P$_0$ after a variable hold time.  Repump light at 707~nm is applied during the second image to depopulate $^3$P$_2$.  We measure a depopulation rate of $7(1) \times 10^{-4}/(E_R s)$, from which we infer a decoherence rate of $4.1(4) \times 10^{-4}/(E_R s)$, in agreement with the predictions of ref.~\cite{hutson2019engineering} and represented by the dashed gray line in the inset of fig.~\ref{fig:coherence} of the main text.  During images used in this work, we infer a 5-9$\%$ chance of depopulating $^3$P$_0$.  Note that atoms that are scattered from $^3$P$_0$ are returned to the ground state by optical pumping for future cycles of the experiment, and that all reported contrasts in this work include this depopulation effect.

\noindent \textbf{Sideband cooling. } We perform three-dimensional resolved-sideband cooling before probing the clock transition.  This requires ``magic'' trapping conditions such that the transition frequency does not depend on the position of the atoms in their traps. To do this, we control the angle between a large ($\sim 30$~G) magnetic field and the polarization of the optical tweezers as described in \cite{norcia2018microscopic}. 
In the tightly confined direction perpendicular to the tweezer axis, along which the clock light is applied, we measure an average of 0.2 quanta of motional excitation using sideband thermometry on the $^1$S$_0$ to $^3$P$_1$ transition.  In the more weakly confined direction along the tweezer axis, the motional sidebands are poorly resolved.  However, from the width of the observed $^3$P$_1$ excitation feature when probed along the axial direction, we bound the mean number of excitations in the axial direction to less than $\sim 2$.  

\begin{table}[!htb]
	\centering
	\begin{tabular}{|r|c|c|} 
		\hline
		\textbf{Set} & \textbf{B~(G)} & \textbf{$ \mathrm{\Omega}$~(Hz)}\\ [0.5ex] 
		\hline
		Fig. 1 & 3.7 & 0.33 \\ 
		Figs. 2a, 2b, 3aI, 4 & 22 & 17\\
		Fig. 3aII & 2.8 & 0.25 \\
		Fig. 3aIII & 1.4 & 0.125\\ [1ex] 
		\hline
	\end{tabular}
	\caption{Summary of applied magnetic fields and clock Rabi frequencies for data used in the main text.}
	\label{tab:params}
\end{table}

\noindent \textbf{Clock interrogation. } We apply a variable magnetic field --- in the range of 1.4 to 22 Gauss for the range of Rabi frequencies presented here (see tab.~\ref{tab:params}) --- along axis of the tweezers to induce the clock transition in $^{88}$Sr \cite{taichenachev2006magnetic, origlia2018towards}.  For this range of magnetic fields, we measure a quadratic Zeeman shift ranging from 0.4~Hz to 90~Hz.  After sideband cooling, we adiabatically ramp from the magic field condition to this condition in 10~ms. At the same time, we ramp down the depth of the tweezers to alleviate dephasing associated with differential light shifts due to the tweezers being at slightly different frequencies. We can then apply resonant clock light to probe this transition (either in a Rabi or Ramsey configuration in this work). At the end of this sequence, we ramp back to a zero-field configuration in 15~ms to image the atoms again. Atoms that have been excited to the clock state appear dark to the imaging and cooling light, and so we can infer the excitation to the clock state from the apparent atom loss. The atoms are then repumped to the ground state through the $^3$P$_0$ to $^3$S$_1$ and $^3$P$_2$ to $^3$S$_1$ transitions (at 679~nm and 707~nm respectively) so that they can be recycled for another round of imaging and clock interrogation without reloading the tweezers. This repeated interrogation also allows us to distinguish between excitation and actual atom loss.

For the relatively shallow tweezers compatible with our longest measured coherence times, we observe an initial contrast of around 80\%.  This contrast improves with tweezer depth up a peak value of 90\% at a tweezer depth of 900~$E_R$, suggesting the presence of residual motional effects, either associated with motional excitation before or during interrogation, or with coupling to the relatively warm direction along the tweezer axis.  Working with shallow traps is desirable in order to minimize sources of decoherence associated with the trap light due to scattering or inhomogeneous light shifts, and to minimize systematic shifts from imperfect Stark shift cancellation.  In principle, the optical pulses used for Ramsey spectroscopy could be applied in deep tweezers, whose depth could then be adiabatically lowered for the dark evolution time, enabling both high contrast and low dephasing.

\noindent \textbf{Clock light. } The clock light is stabilized to a silicon cavity operated at cryogenic temperatures, as described in \cite{oelker2019optical}.  Light is delivered to the tweezer experiment via a series of injection locks and noise-canceled fibers.  The final 2~m of fiber before the atoms, and a roughly 50~cm long free-space path to the atoms are not actively canceled.  

\noindent \textbf{Allan deviation and quantum projection noise prediction.}
We measure Allan deviation by performing Ramsey spectroscopy with a dark evolution time chosen for maximum sensitivity to frequency, where the average excitation fraction is half way between its maximum and minimum values.  We then use the evolution time and fitted contrast to estimate the conversion between excitation fraction and frequency.  We terminate the measurement before the average excitation fraction drifts significantly from the point of maximal sensitivity, so we bound the error on this conversion factor to 5\%.  In computing the quantum projection noise level, we use the measured average atom number of 4.8 and apply a correction factor of 1.07, which represents excess noise arising from the fact that the atom number fluctuates between trials.  This value was determined through simulation, assuming that the number of atoms present in a given trial is sampled from a binomial distribution with 10 trials and a success probability of 0.48.  For these conditions, we estimate Dick noise to be $3 \times 10^{-17}(\tau/s)^{-1/2}$, which when added in quadrature to the predicted quantum projection noise leads to a correction well below 1\%.  

In order to reach the stability of current state-of-the-art optical lattice clocks ($4.8 \times 10^{-17}(\tau/s)^{-1/2}$\cite{oelker2019optical}), we could use 2~second long Ramsey measurements and approximately 150 atoms.  This estimate assumes that the contrast with this higher atom number is similar to our current measured value for 2~second Ramsey evolution times (53\%).  We assume a single interrogation per ensemble of atoms, and a total dead-time of 390~ms between interrogations.  This estimate accounts for laser noise estimated from a noise model for our silicon-cavity-stabilized laser system.  

\begin{figure}[!htb]
	\includegraphics[width = \linewidth]{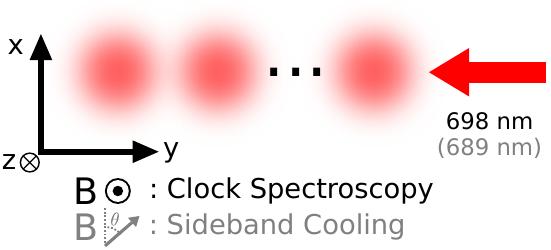}
	\caption{Schematic of laser and magnetic field orientations. In the coordinates shown here, the tweezer light is linearly polarized along the x-axis, and the clock light (698 nm) is linearly polarized along the z-axis. For sideband cooling, we tune our magnetic field to a magic angle of $\theta = 49{\degree}$ relative to the tweezer polarization. Clock light propagates along the y-axis (red arrow) and three 689 nm cooling beams are applied with projections along all three coordinate axes. }
	\label{fig:beams}
\end{figure}

\subsection{Supplementary Text}

\begin{figure}[!htb]
	\includegraphics[width = \linewidth]{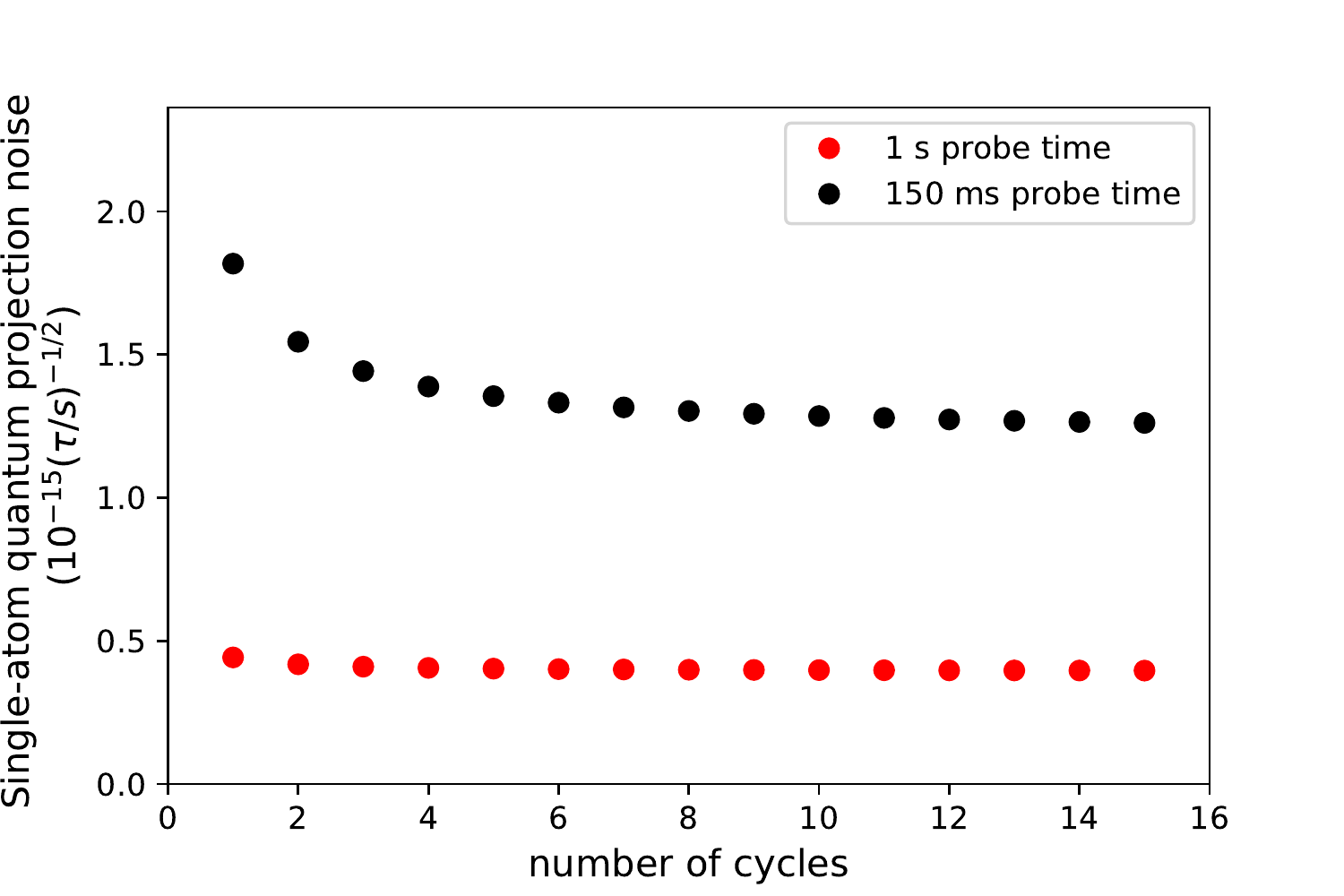}
	\includegraphics[width = \linewidth]{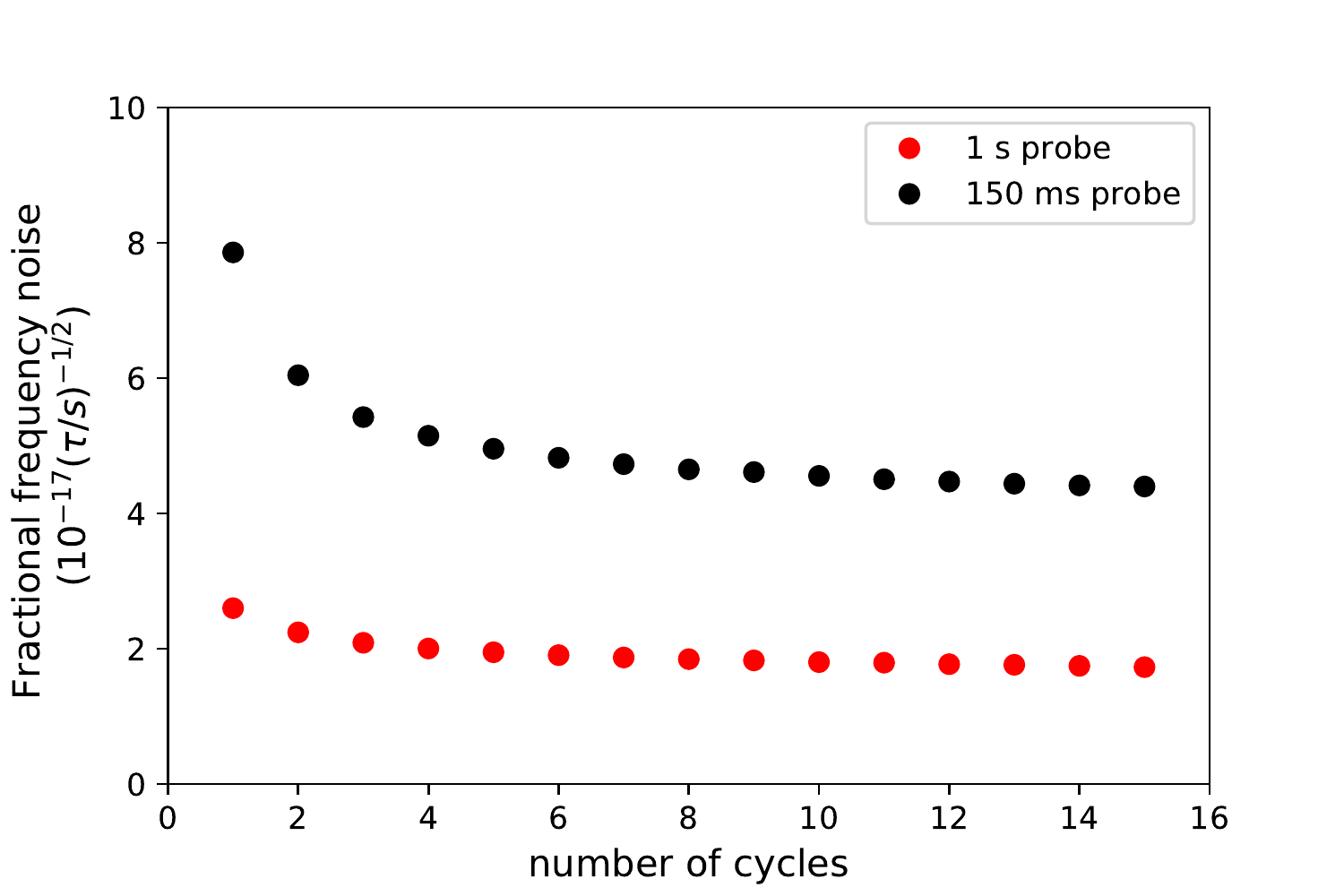}
	\caption{Predicted quantum projection noise limited instability for a single atom (upper) and Dick noise aliasing (lower) versus number of interrogation cycles per ensemble of atoms loaded.  Here we have used an atom load time of 300~ms, a dead-time per interrogation cycle of 90~ms, and a Ramsey sequence with short pulses and a free-evolution time of either 150~ms or 1~second.  }
	\label{fig:qpndick}
\end{figure}

\begin{figure}[!htb]
	\includegraphics[width = \linewidth]{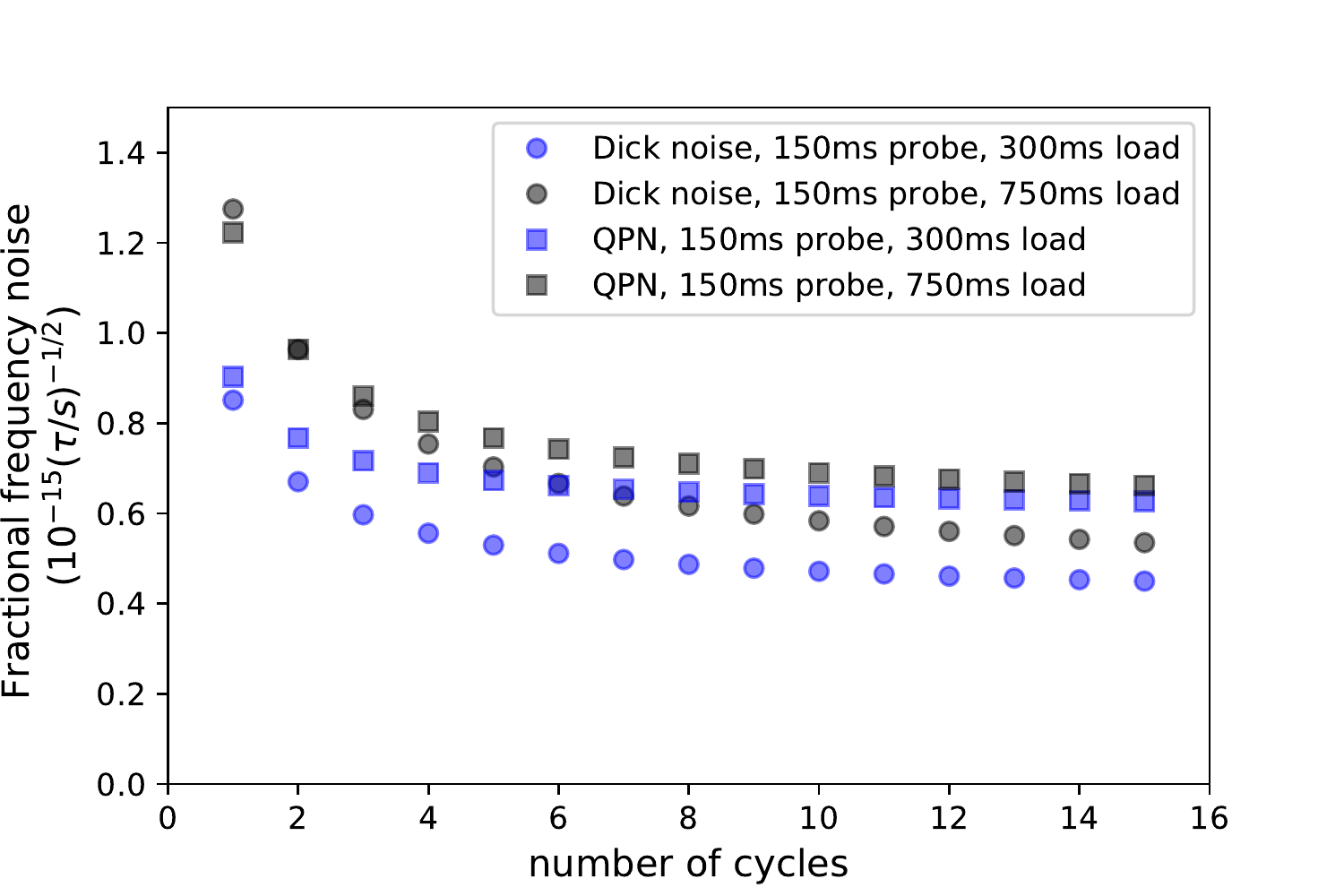}
	\caption{Predicted Dick effect and quantum projection noise (QPN) for laser properties representative of a portable device.  QPN is calculated assuming an average atom number of 5.  Load times of 300~ms and 750~ms are shown.  }
	\label{fig:portable}
\end{figure}

\noindent \textbf{Repeated interrogation and Dick effect noise.}
Operating with high duty cycle by interrogating the same atoms multiple times is advantageous both for rapidly acquiring statistics to average down atom shot noise, and for mitigating noise aliasing of the clock laser, known as the Dick effect.  We show predictions for these sources of instability based on our experimental parameters for different numbers of cycles per repetition, and either short, 150~ms interrogation times or long, 1~second interrogation times (fig.~\ref{fig:qpndick}).  
For all imaging durations, both of these sources of instability are improved by interrogating multiple times.  

For the quantum projection noise (fig.~\ref{fig:qpndick}, upper), the improvement is simply due to the fact that one can make more measurements in a given amount of time, allowing one to average down atom shot noise faster.  The impact of repeated interrogation is larger for short interrogation times, as the total repetition time per measurement is dominated by atom loading for single interrogations, and balanced more evenly between interrogation and imaging for large cycle numbers.  For long probe times, however, the total repetition time is dominated by the interrogation period for either single or repeated imaging, and so the effect is small.  

In principle, the advantages of improved duty cycle will be offset by the atom loss after some number of cycles per ensemble loading.  However, in our experiment the losses are low enough that this becomes significant only after large numbers of repetitions, where the wins associated with repeated imaging have already saturated.  In practice we find that the optimal number of cycles is set more by practical considerations such as the length and density of our datasets.  

Dick noise aliasing, which arises from the periodically modulated sensitivity of a clock sequence to changes in laser frequency when dead time is present, is mitigated by the improved duty cycle of repeated imaging (fig.~\ref{fig:qpndick}, lower).  While such noise aliasing has been a persistent limitation for optical lattice clocks, our laser system has a low noise floor of $\sigma_y = 4.6 \times 10^{-17}$ \cite{oelker2019optical} (expressed as a fractional Allan deviation) and we operate with a small number of atoms so for our current conditions this noise is not a limiting factor for any operating conditions. 

However, Dick noise aliasing remains an important limitation for portable optical lattice clocks \cite{grotti2018geodesy}, where technical constraints necessitate lower performance laser systems.  This increases the noise associated with the Dick effect both because there is more laser noise to alias, and because a noisier laser limits the duration over which the atoms can be interrogated, and thus the duty cycles achieved.  

In fig.~\ref{fig:portable}, we present calculations of Dick noise and quantum projection noise (QPN), assuming parameters compatible with portable operation.  We have assumed a laser with a flicker noise floor $\sigma_y = 4 \times 10^{-16}$ \cite{koller2017transportable}, and a white noise contribution of $7 \times 10^{-16}/\sqrt{Hz}$.  QPN is calculated for our current atom number $N=5$, and for simplicity assumes full contrast.  Here, we use a Ramsey interrogation sequence with 150~ms dark evolution time and 1.5~ms $\pi$ pulses, 90~ms of dead time per cycle, and either 300 or 750~ms to load a new ensemble of atoms.  These load times reflect our current value, and a somewhat longer value reflecting constraints on atomic flux and laser power that may be associated with a portable system \cite{koller2017transportable}.  With these assumptions, the contributions to total noise from Dick noise aliasing and QPN are comparable, even at our current atom number, and both improve with repeated imaging.

\end{document}